\documentstyle[psfig]{mn}

\title{Primordial feature at the scale of superclusters of galaxies}
\author[Mirt Gramann and Gert H\"utsi]
{Mirt Gramann and Gert H\"utsi
   \\
   Tartu Observatory,
       T\~oravere 61602, Estonia}
\begin{document}
\maketitle

\let\sec=\section
\let\ssec=\subsection
\let\sssec=\subsubsection

\def\kms{\;{\rm km\,s^{-1}}}
\def\kmsmpc{\;{\rm km\,s^{-1}\,Mpc^{-1}}}
\def\hompc{\,h\,{\rm Mpc}^{-1}}
\def\mpcoh{\,h^{-1}\,{\rm Mpc}}
\def\mpc3h{\,h^{3}\,{\rm Mpc^{-3}}}

\begin{abstract}
We investigate a spatially-flat cold dark matter model (with the matter
density parameter $\Omega_m=0.3$) with a primordial feature in the
initial power spectrum. We assume that there is a bump in the
power spectrum of density fluctuations at wavelengths
$\lambda \sim 30-60h^{-1}$Mpc,
which correspond to the scale of superclusters of galaxies. There
are indications for such a feature in the power spectra derived
from redshift surveys and also in the power spectra derived from
peculiar velocities of galaxies. We study the mass function of
clusters of galaxies, the power spectrum of the CMB temperature
fluctuations, the rms bulk velocity and the rms peculiar velocity of clusters
of galaxies. The baryon density is assumed to be
consistent with the BBN value. We show that with an appropriately
chosen feature in the power spectrum of density fluctuations at the
scale of superclusters, the mass function of clusters, the CMB
power spectrum, the rms bulk velocity and the rms peculiar velocity of
clusters are in good agreement with the observed data.
\end{abstract}

\begin{keywords}
galaxies: clusters: general -- cosmology: theory -- dark matter --
large-scale structure of Universe -- cosmic microwave background.
\end{keywords}

\sec{INTRODUCTION}

The spatially-flat cold dark matter (CDM) model (matter density
parameter $\Omega_m=0.3$, flatness being restored by a
contribution from a cosmological constant $\Omega_{\Lambda}=0.7$)
with a scale-invariant initial conditions has been a standard
model in cosmology over last five years (see e.g. Ostriker \&
Steinhardt 1995). This model successfully explains many large- and
small-scale structure observations including the
mass function and the peculiar velocities of clusters of
galaxies. The flat cosmological model with the matter density
parameter $\Omega_m=0.3$ is also consistent with the observations
of Type Ia supernovae at redshift $z \sim 1$ (Perlmutter at al.
1999; Riess et al. 1998).

In this paper we investigate the CDM model with a primordial feature
in the initial power spectrum of density fluctuations. Adams, Ross \&
Sarkar (1997) have noted that according to our current understanding of
the unification of fundamental interactions, there should have been phase
transitions associated with spontaneous symmetry breaking during the
inflationary era. This may have resulted in the breaking of scale-invariance
of the initial power spectrum. Chung et al. (2000) studied an alternative
mechanism that can alter classical motion of the inflaton and produce
features in the initial power spectrum. They showed that if the inflaton
is coupled to a massive particle, resonant
production of the particle during inflation modifies the evolution of
the inflaton, and may leave an imprint in the initial power spectrum.
The spectral features in the initial spectrum may also be generated if
the inflaton evolves through a kink in its potential (Starobinsky 1992;
Lesgourgues, Polarski \& Starobinsky 1998; Gramann \& H\"utsi 2000).

A number of different non-scale invariant initial conditions have been
recently used to analyze the cosmic microwave background (CMB) data
(see e.g. Kanasawa et al. 2000; Barriga et al. 2000;
Atrio-Barandela et al. 2000; Griffiths, Silk \& Zaroubi 2000;
Hannestad, Hansen \& Villante
2000; Wang \& Mathews 2000). Griffiths, Silk \& Zaroubi (2000) and
Hannestad, Hansen \& Villante (2000) showed that the CMB data favour
a bump-like feature in the power spectrum around a scale of
$k=0.004h$Mpc$^{-1}$. Barriga et al. (2000) studied the step-like spectral
feature in the range $k\sim (0.06-0.6)h$Mpc$^{-1}$ and found that
such a spectral break enables a good fit to both the APM and CMB data.
Atrio-Barandela et al. (2000) investigated the temperature power spectrum
in the CDM models, where the power spectrum of density fluctuations
at $z\sim 10^3$ was in the form $P(k) \sim k^{-1.9}$ at wavenumbers
$k>0.05h$Mpc$^{-1}$. This power spectrum of density fluctuations was
derived by Einasto et al. (1999) by analyzing different observed power
spectra of galaxies and clusters of galaxies. Atrio-Barandela et al. (2000)
found that this form of the power spectrum of density fluctuations is
consistent with the recent CMB data. However, in this paper we examine
the mass function of clusters of galaxies in the same model and find that
for $\Omega_m=0.3$, the number density of clusters is
significantly smaller than observed.

Suhhonenko \& Gramann (1999, hereafter SG) studied properties of
clusters of galaxies in two cosmological models which rely on the
observed power spectra of the distribution of galaxies. In the first
model (hereafter model 1), the power spectrum of density fluctuations at
$z\sim 10^3$ was in the form $P(k) \sim k^{-2}$ at wavenumbers
$k>0.05h$Mpc$^{-1}$. In the second model (hereafter model 2), the power
spectrum contained a feature (bump) at wavenumbers
$k \sim 0.1-0.2 h$Mpc$^{-1}$ ($\lambda
\sim 30 - 60h^{-1}$Mpc), which correspond to the scale of
superclusters (see e.g. Einasto et al. 1997). SG examined the mass
function, peculiar velocities, the power spectrum and the correlation
function of clusters in both models and found that in many aspects the power
spectrum of density fluctuations in  model 2 fits the observed
data better than the simple power-law model 1. This study
suggested that probably at wavenumbers $k\sim 0.05-0.2 h$Mpc$^{-1}$, the
power spectrum of density fluctuations is not a featureless simple power law.

Fig.~1 shows the power spectrum of density fluctuations in the
CDM model examined in this paper (see eq. (5) below). We assume
that there is a feature (bump) in the power spectrum at wavenumbers
$k=0.1-0.2h^{-1}$ Mpc. The power spectrum in the CDM model with a
scale-invariant initial power spectrum is also plotted. It is assumed
that the density parameter $\Omega_m=0.3$ and the normalized Hubble
constant $h=0.65$. For comparison, we show in Fig.~1 the observed
power spectra derived from the distribution of galaxies in the
APM, Stromlo-APM and Durham/UKST surveys (Baugh \&
Efstathiou 1993; Tadros \& Efstathiou 1996; Hoyle et al. 1999).
For the Stromlo-APM and Durham/UKST surveys, we present
estimates for the flux-limited sample with $P(k)=8000h^{-3}$ Mpc$^3$ in
the weighting function (see Tadros \& Efstathiou 1996; Hoyle et al.
1999 for details). There are indications for a similar bump
in the power spectrum derived from the Stromlo-APM survey. On the other
hand, there is no similar feature in the
APM and Durham/UKST power spectrum. Hoyle et al. (1999) analyzed
the power spectrum for different volume-limited and flux-limited samples
drawn from the Durham/UKST redshift survey. There are indications for a
similar bump in the power spectrum measured in a volume limited sample
with $z_{max}=0.08$ (see Hoyle et al. (1999) for details).

We also examined the power spectrum derived from the distribution of
galaxies in the SSRS2+CfA2 redshift survey (da Costa et al. 1994). There
is a feature in the SSRS2+CfA2 power spectrum at wavenumbers
$k=0.1-0.2h^{-1}$ Mpc. Silberman et al. (2001) studied the
power spectrum of peculiar velocities of galaxies and found an
indication for a wiggle in the power spectrum: an excess near
$k\sim 0.05h$ Mpc$^{-1}$ and a deficiency at $k\sim 0.1h$
Mpc$^{-1}$. This wiggle in the power spectrum is similar to the
spectral feature studied in this paper. Most recently, there are
indications for such a feature in the preliminary power spectrum
derived from part of the 2dF redshift survey (Percival et al. 2001).

\begin{figure}
\centering
\leavevmode
\psfig{file=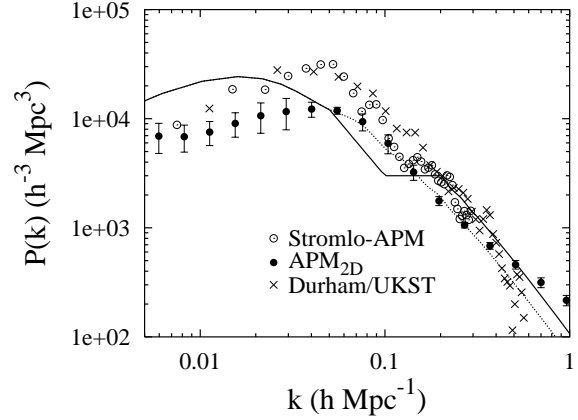,width=8cm}

\caption {The power spectrum of density fluctuations in the CDM
model examined in this paper (solid line) and in the standard CDM
model with a scale-invariant initial power spectrum (dotted line). In
the models studied, $\Omega_m=0.3$ and $h=0.65$. Filled circles, open
circles and crosses show the power spectrum of the galaxy
distribution in the APM, Stromlo-APM and Durham/UKST
surveys, respectively. For clarity, error bars are only shown for the
APM data.}
\end{figure}

SG studied the power spectrum of clusters using N-body simulations
and showed that the power spectrum of clusters in model 2 is in
good agreement with the observed power spectrum of the APM clusters
determined by Tadros, Efstathiou \& Dalton (1998). SG investigated also
the relation between the power spectrum of clusters and the power
spectrum of matter fluctuations and found that in this model the relation
between the cluster power spectrum and matter power spectrum is not
linear at wavenumbers $k>0.1h$Mpc$^{-1}$ (see also Gramann \& Suhhonenko 1999).

In this paper we investigate the mass function of clusters of
galaxies and the temperature power spectrum in the model with a
bump in the power spectrum of density fluctuations at the scale
of superclusters of galaxies. We also study the rms bulk velocity
of galaxies and the rms peculiar velocity of clusters of galaxies
in this model. The results are compared with observations. We
examine the flat cosmological model with the density parameter
$\Omega_m=0.3$, the baryon density $\Omega_b h^2=0.019$ and the
normalized Hubble constant $h=0.65$ and $h=0.70$. These values
are in agreement with measurements of the density parameter (e.g.
Bahcall et al. 1999), with measurements of the baryon density
from abundances of light elements (O'Meara et al. 2001; Tytler et al.
2000) and with measurements of the Hubble constant using various
distance indicators (Freedman et al. 2000; see also Parodi et al.
2000). The Hubble constant is given as $H_0=100h$ km s$^{-1}$
Mpc$^{-1}$. To restore the spatial flatness in the low density
model, we assume a contribution from a cosmological constant
$\Omega_{\Lambda}=0.7$.

To study the mass function of clusters we use the Press-Schechter
(Press-Schechter 1974, hereafter PS) approximation. The
transfer functions $T(k)$ and the temperature power spectra are
calculated using the
fast Boltzmann code CMBFAST developed by Seljak \& Zaldarriaga (1996).
The code CMBFAST has been modified to incorporate a primordial feature
in the initial power spectrum. We assume that the
initial fluctuations are adiabatic and that the initial density fluctuation
field is a Gaussian field. In this case, the power spectrum provides a
complete statistical description of the field.

This paper is organized as follows. In Section 2 we study the mass
function of clusters of galaxies and temperature power spectrum in our
model, and compare the results with observations. In Section 3 we
examine peculiar velocities of galaxies and clusters of galaxies.
Discussion and summary are presented in Section 4.

\sec{THE MASS FUNCTION OF CLUSTERS AND CMB ANISOTROPIES}

Let us first consider the CDM model, where the power spectrum of density
fluctuations at $z\sim 10^3$ is in the form
$$
P(k)=AkS(k)T^{2}(k)=\cases{AkT^{2}(k), & if $k<k_0$;\cr
                         P(k_0)(k/k_0) ^{-1.9}, & if $k>k_0$, \cr}
\eqno(1)
$$
where $k_0=0.05h$ Mpc$^{-1}$. Here, the initial power spectrum is
defined as $P_{in}(k)=AkS(k)$ and $T(k)$ is the transfer
function, which describes the modification of the initial power
spectrum during the era of radiation domination. The function
$S(k)$ describes the deviation of the initial power spectrum from
a scale invariant form $P(k) \sim k$. The normalization constant
$A$ is determined by the large-scale CMB anisotropy. This form of
the power spectrum at wavenumbers $k>0.05h$Mpc$^{-1}$ was
derived by Einasto et al. (1999) by analyzing different observed
power spectra of galaxies and clusters of galaxies.
We denote the CDM model, were the power spectrum is in the form (1),
as model 1.

Fig.~2a shows the power spectrum of density fluctuations in model 1
for $h=0.65$ and $h=0.70$. We also show the power spectrum in
the CDM model with a scale-invariant
spectrum ($S(k)\equiv 1$). In comparison with the standard model,
the power spectrum of density fluctuations in model 1 is depressed
at wavenumbers $k>0.05h$Mpc$^{-1}$. A similar break in the power
spectrum of density fluctuations was analyzed by
Atrio-Barandela et al. (2000).

\begin{figure}
\centering
\leavevmode
\psfig{file=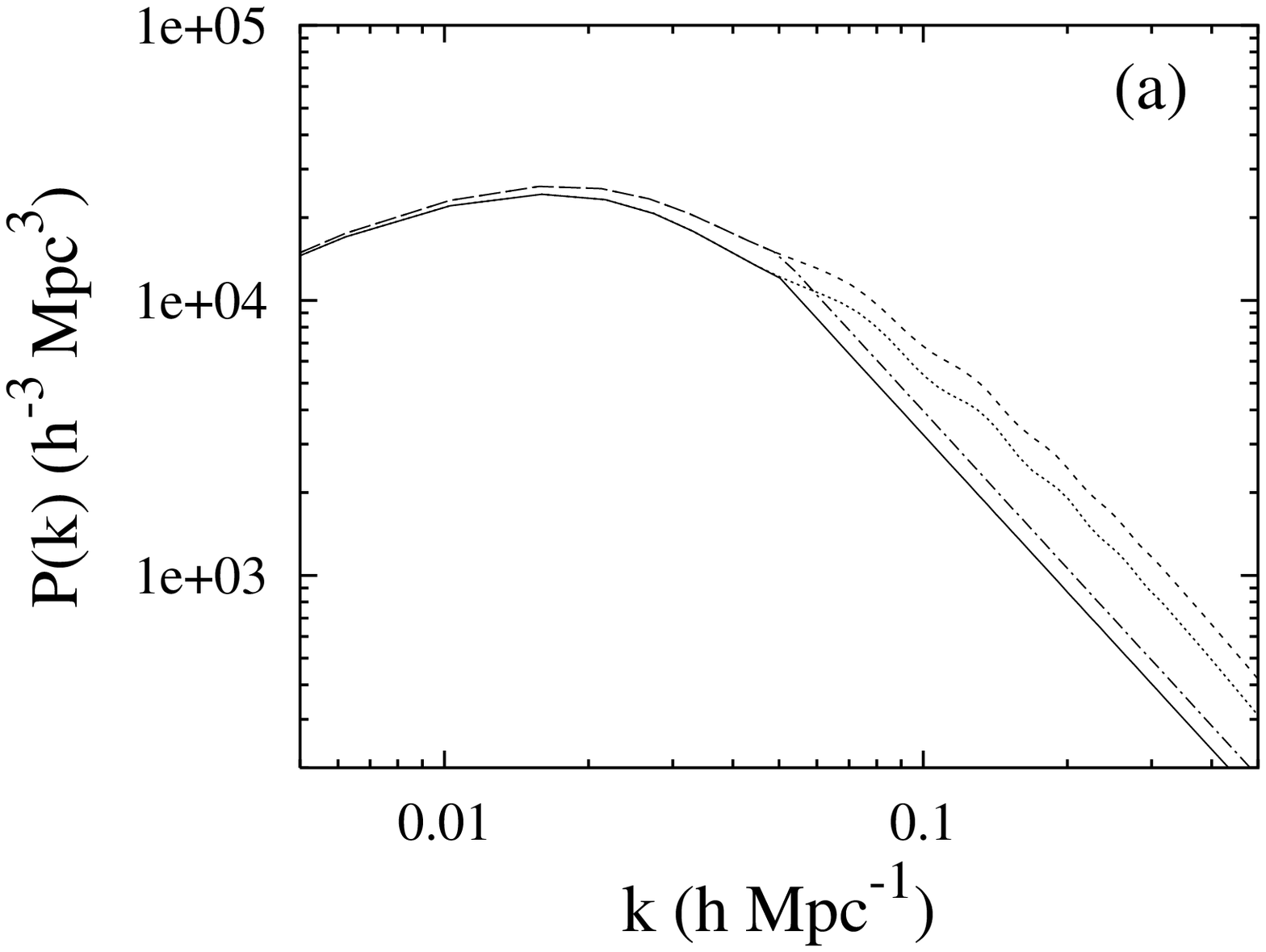,width=8cm}
\psfig{file=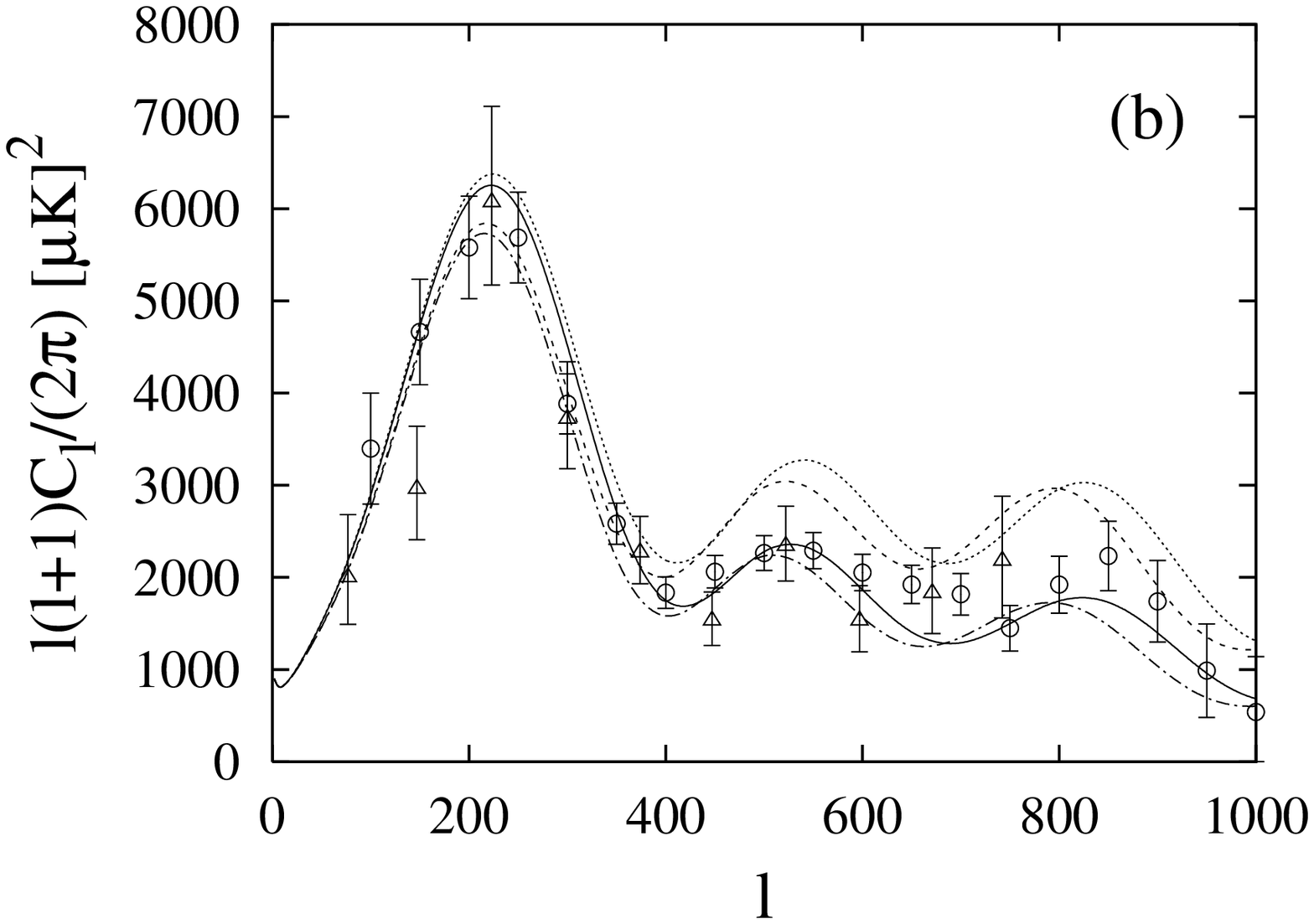,width=8cm}
\psfig{file=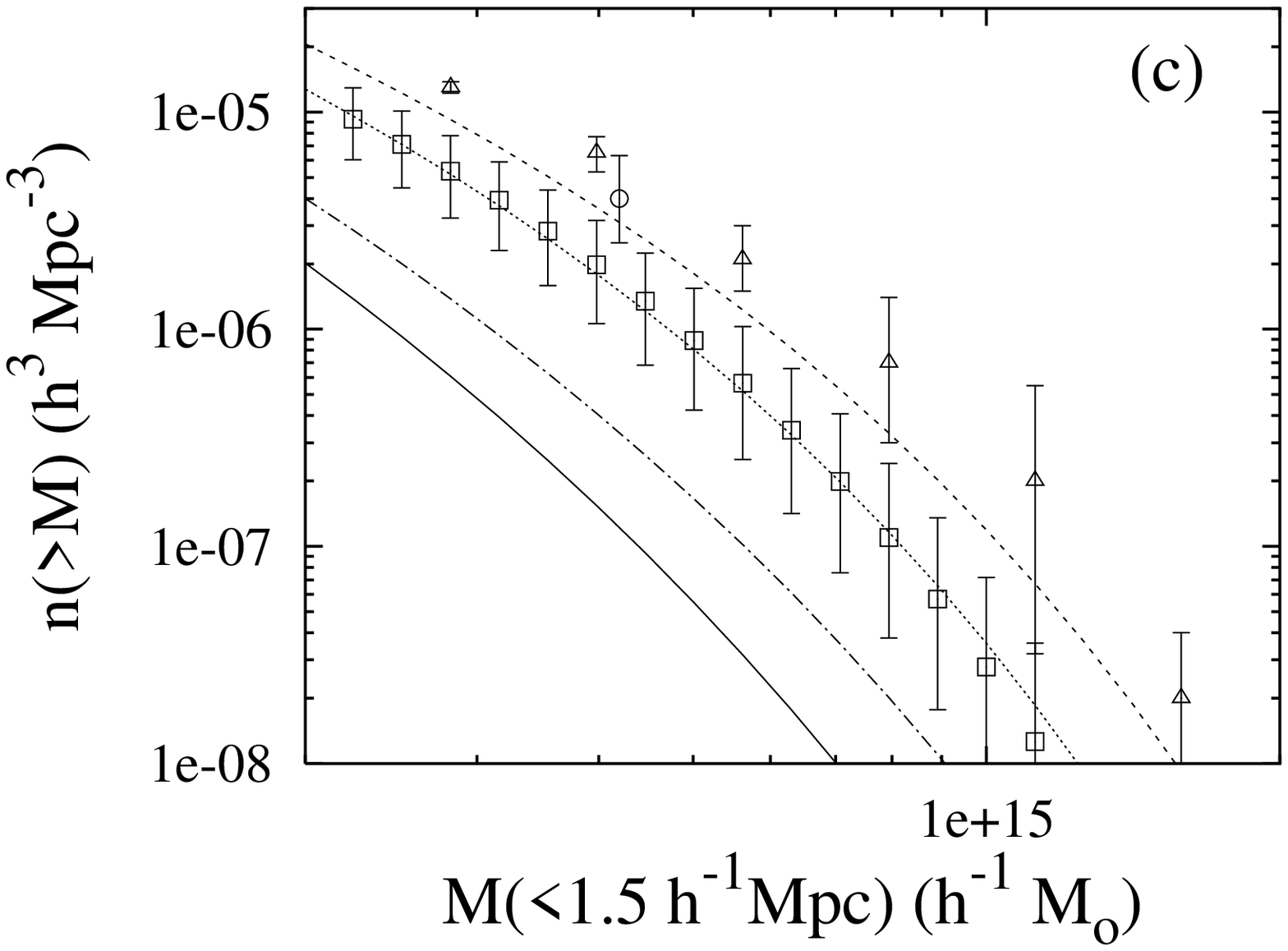,width=8cm}

 \caption{(a) The power spectrum of density fluctuations in
 model 1 for $h=0.65$ (solid
line) and for $h=0.70$ (dot-dashed line). The dotted line and
dashed line show the power spectrum in the standard CDM model for
$h=0.65$ and $h=0.70$, respectively. (b) CMB power spectrum in
the same models. The data are from the Boomerang (circles) and
Maxima-1 (triangles) experiments. (c) The cluster mass function
in the same models. Squares and triangles show the cluster
mass function derived by Bahcall \& Cen (1993) and Girardi et al.
(1998), respectively. The circle represents the result
obtained by White et al. (1993)}

\end{figure}

To calculate the angular power spectrum of the CMB temperature fluctuations,
we used the code CMBFAST, which was modified to incorporate the
function $S(k)$ in the initial power spectrum. We examined the models
with no reionization (optical depth $\tau=0$).
Fig.~2b shows the CMB power spectrum, $\Delta T_l^2=l(l+1)C_l/2\pi$,
predicted in model 1. Here $C_l= \langle a^2_{lm} \rangle$
and $a_{lm}$ are the coefficients of the spherical harmonic
decomposition of the CMB temperature field: $\Delta
T(\theta,\varphi)=\sum a_{lm} Y_{lm}(\theta,\varphi)$.
Fig.~2b shows also the temperature power spectrum predicted in the
standard CDM model. We see that lowering the amplitude of the
power spectrum at wavenumbers $k>0.05h^{-1}$Mpc, lowers also the
amplitude of the temperature power spectrum at multipoles $l>400$.
The height of the first peak is different in the models in Fig.~2b due
to the change in Hubble's constant.

Fig.~2b shows also the CMB power spectrum derived from the
Boomerang (Netterfield et al. 2001) and from the Maxima-1
(Hanany et al. 2000) experiments. We see that the temperature power
spectrum in model 1 is consistent with the observed
temperature power spectrum. The amplitude of the second acoustic peak
at $l\sim 500$ in the observed temperature power spectrum is smaller
than that predicted in the standard CDM model with $\Omega_m=0.3$.
However, in the analyzes we have assumed that the
spectral index $n=1$ and the optical depth $\tau=0$.
Also, both CMB datasets have a
calibration uncertainty and a beam uncertainty that are not
included in the errors plotted in Fig.~2b.
Netterfield et al. (2001) analyzed the CMB power spectrum in the
standard CDM model in more detail and showed that this model is
consistent with the observed CMB data, once the parameters $n\neq 1$
and $\tau\neq 0$, and beam and calibration uncertainties are taken into
account.

To study the mass function of clusters we use the Press-Schechter
(1974, PS) approximation. The PS mass function has been compared
with N-body simulations (Efstathiou et al. 1988; White,
Efstathiou \& Frenk 1993; Lacey \& Cole 1994; Eke, Cole \& Frenk 1996)
and has been shown to provide an accurate description of
the abundance of virialized cluster-size halos. In the PS approximation
the number density of clusters with the mass between $M$ and $M+dM$ is
given by
$$
n(M) dM = - \sqrt{{2 \over \pi}} {\rho_b \over M}
{\delta_t \over \sigma^2(M)} {d\sigma(M)\over dM}
\exp \left[-{\delta_t^2 \over 2\sigma^2(M)}\right]  dM .
\eqno(2)
$$
Here $\rho_b$ is the mean background density and $\delta_t$ is the
linear theory overdensity for a uniform spherical fluctuation which is
now collapsing; $\delta_t=1.675$ for $\Omega_0=0.3$ (Eke et al. 1996).
The function $\sigma(M)$ is the rms linear density fluctuation at the
mass scale $M$. We will use the top-hat window function to describe halos.
For the top-hat window, the mass $M$ is related to the window radius
$R$ as $M=4\pi\rho_b R^3/3$. In this case, the number density of clusters
of mass larger than $M$ can be expressed as
$$
n_{cl}(>M) = \, {\int_M^{\infty}} n(M') dM' =
$$
$$
= - {3 \over (2\pi)^{3/2}} {\int_R^{\infty}} { \delta_t \over \sigma^2(r)}
{d\sigma(r)\over dr} \exp\left[-{\delta_t^2 \over 2\sigma^2(r)}\right]
{dr \over r^3}  .
\eqno(3)
$$

Fig.~2c shows the cluster mass function predicted in model 1 for
$h=0.65$ and $h=0.70$. The mass function in the standard CDM model is
also plotted. We investigated the cluster masses within a $1.5h^{-1}$ Mpc
radius sphere around the cluster center. This mass
$M_{1.5}$ is related to the window radius $R$ as
$$
R=8.43 \Omega_0^{0.2\alpha \over 3-\alpha} \left[{M_{1.5} \over 6.99
\times 10^{14} \Omega_0 h^{-1} M_{\odot}}\right]
^{1 \over 3-\alpha} (h^{-1} {\rm Mpc})  .
\eqno(4)
$$
Here the parameter $\alpha$ describes the cluster mass profile,
$M(r) \sim r^\alpha$, at radii $r\sim 1.5h^{-1}$ Mpc. Numerical
simulations and observations of clusters indicate that the parameter
$\alpha \approx 0.6-0.7$ for most of clusters  (e.g. Navarro, Frenk \&
White 1995; Carlberg, Yee \& Ellingson 1997). In this paper we use the
value $\alpha=0.65$.

Fig.~2c also shows the mass function of clusters derived by Bahcall \&
Cen (1993, hereafter BC) and by Girardi et al. (1998, hereafter G98).
BC used both optical and X-ray observed properties of clusters to determine
the mass function of clusters. The function was extended towards the faint
end using small groups of galaxies. G98 determined the mass function
of clusters by using virial mass estimates for 152 nearby
Abell-ACO clusters including the ENACS data (Katgert et al. 1998). The
mass function derived by G98 is somewhat larger than the mass function
derived by BC, the difference being larger at larger masses (see Fig.~2c).
Reiprich \& Bohringer (1999) determined the cluster mass function using
X-ray flux-limited sample from ROSAT All-Sky survey. They determined
the masses for different outer radii of the clusters and for a radius
$r=1.5h^{-1}$ Mpc their mass function agrees with that determined by
BC.

Let us consider the amplitude of the mass function of galaxy clusters at
$M_{1.5} = 4\cdot 10^{14} h^{-1} M_{\odot}$. For this mass, the cluster
abundances derived by BC and G98 are $n(>M)=(2.0 \pm 1.1)\cdot 10^{-6}
h^3$Mpc$^{-3}$ and $n(>M)=(6.3\pm 1.2)\cdot 10^{-6} h^3$Mpc$^{-3}$,
respectively. By analysing X-ray properties of clusters, White, Efstathiou
\& Frenk (1993) found that the number density of clusters with the mass
$M_{1.5} \approx 4.2 \cdot 10^{14} h^{-1} M_{\odot}$ is
$n(>M)=4 \cdot 10^{-6} h^3$Mpc$^{-3}$.

Fig.~2c shows that the cluster mass function in the standard CDM
model is in good agreement with the observed data. But the number
density of clusters in model 1 is substantially
lower than observed; for the mass $M_{1.5}=4\cdot 10^{14} h^{-1}
M_{\odot}$, the cluster abundance $n(>M)=1.5 \cdot 10^{-7} h^3$Mpc$^{-3}$
and $n(>M)=4.1 \cdot 10^{-7} h^3$Mpc$^{-3}$ for
$h=0.65$ and $h=0.70$, respectively.

In model 1, the amplitude of the power spectrum of density
fluctuations is depressed with respect to the standard CDM model for
$k>0.05h$Mpc$^{-1}$.
Lowering the amplitude of the power spectrum of density fluctuations at
wavenumbers $k>0.05h$ Mpc$^{-1}$ lowers also the CMB
power spectrum at multipoles $l>400$. However, lowering the amplitude of
the power spectrum of density fluctuations lowers also the cluster mass
function, and as a result in model 1 the number density of clusters is smaller
than observed. One possibility to get rid of the last effect is to consider
a bump in the power spectrum of density fluctuations at wavenumbers
$k\sim 0.1-0.2h^{-1}$Mpc. The cluster mass function for masses
$M \sim 10^{14}-10^{15} h^{-1}M_{\odot}$ is sensitive to the amplitude of
the power spectrum at wavenumbers $k\sim 0.2h^{-1}$Mpc, while the
temperature anisotropy at the second acoustic peak is sensitive to the
amplitude of the power spectrum at wavenumbers $k \sim 0.05-0.1h^{-1}$Mpc.

Thus let us now study a CDM model, where the power spectrum of
density fluctuations at $z\sim 10^3$ contains a specific feature at
wavenumbers $k\sim 0.1-0.2h$Mpc$^{-1}$ ($\lambda \sim 30 -
60h^{-1}$Mpc) which correspond to the scale of superclusters:
$$
P(k)=AkS(k)T^{2}(k)=\cases{AkT^{2}(k), & if $k<k_0$;\cr
                           P(k_0)(k/k_0)^{m}, & if $k_0<k<k_1$;\cr
                           P(k_1), & if $k_1<k<k_2$; \cr
                           BkT^{2}(k), & if $k>k_2$, \cr}
\eqno(5)
$$
where $k_0=0.05h$Mpc$^{-1}$, the spectral index
$m=\log{[P(k_1)/P(k_0)]}/\log[{k_1/k_0}]$ and
$B=P(k_1)/(k_{2}T^{2}(k_2))$. This form of the power spectrum
contains three free parameters, which describe the bump in the power
spectrum: $k_1$, $k_2$ and $P(k_1)$. The parameter $k_1$ determines the
beginning of the bump, the parameter $k_2$ - the end of the bump, and the
parameter $P(k_1)$ - the amplitude of the power spectrum for the bump.
In this paper we examine the models where $k_1=0.1h$Mpc$^{-1}$,
$k_2=0.2h$Mpc$^{-1}$ and $P(k_1)=2500-3500h^{-3}$Mpc$^3$.
We denote the CDM model, where the power spectrum is in the form (5), as
model 2.

\begin{figure}
\centering \leavevmode \psfig{file=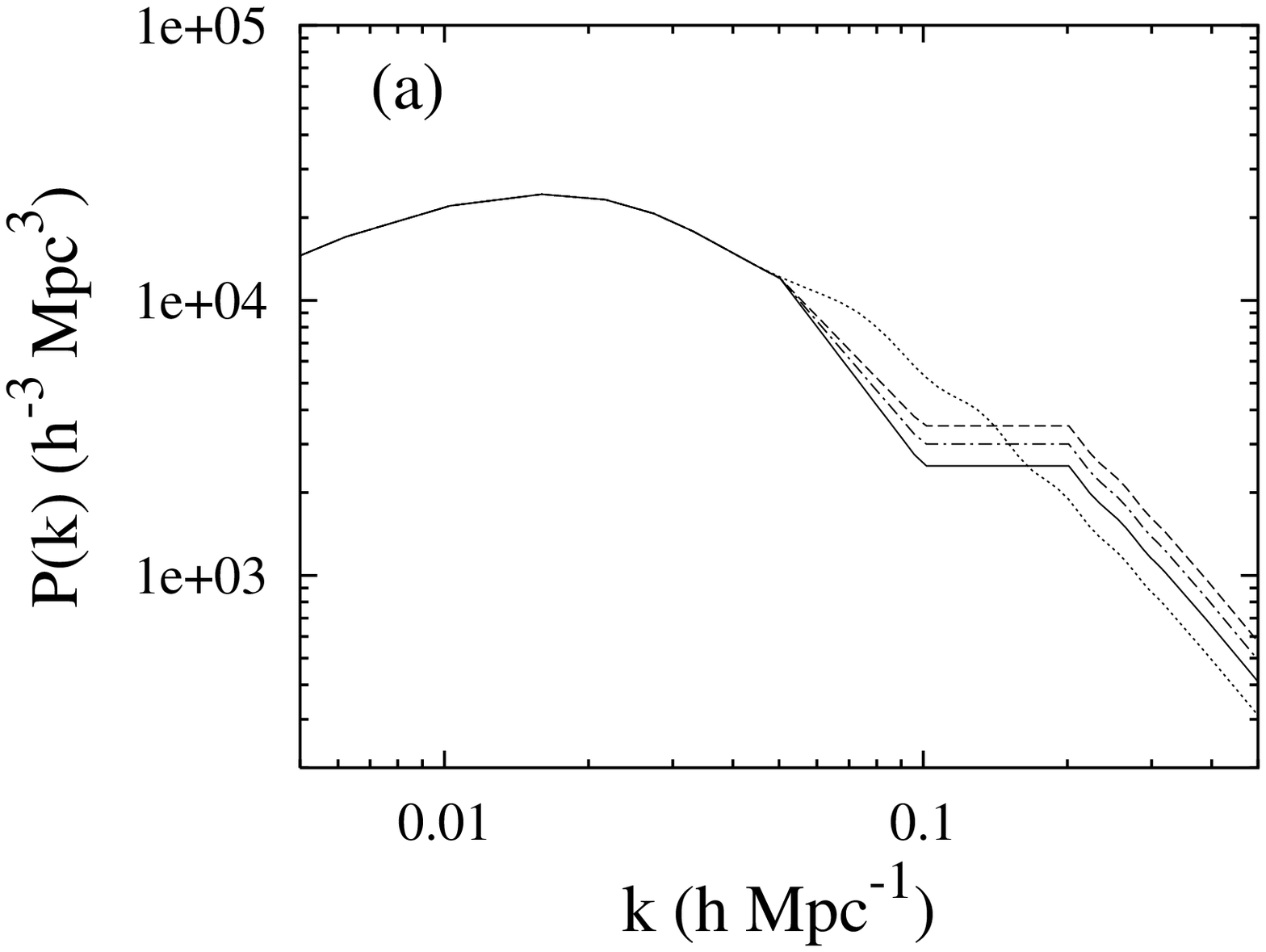,width=8cm}
\psfig{file=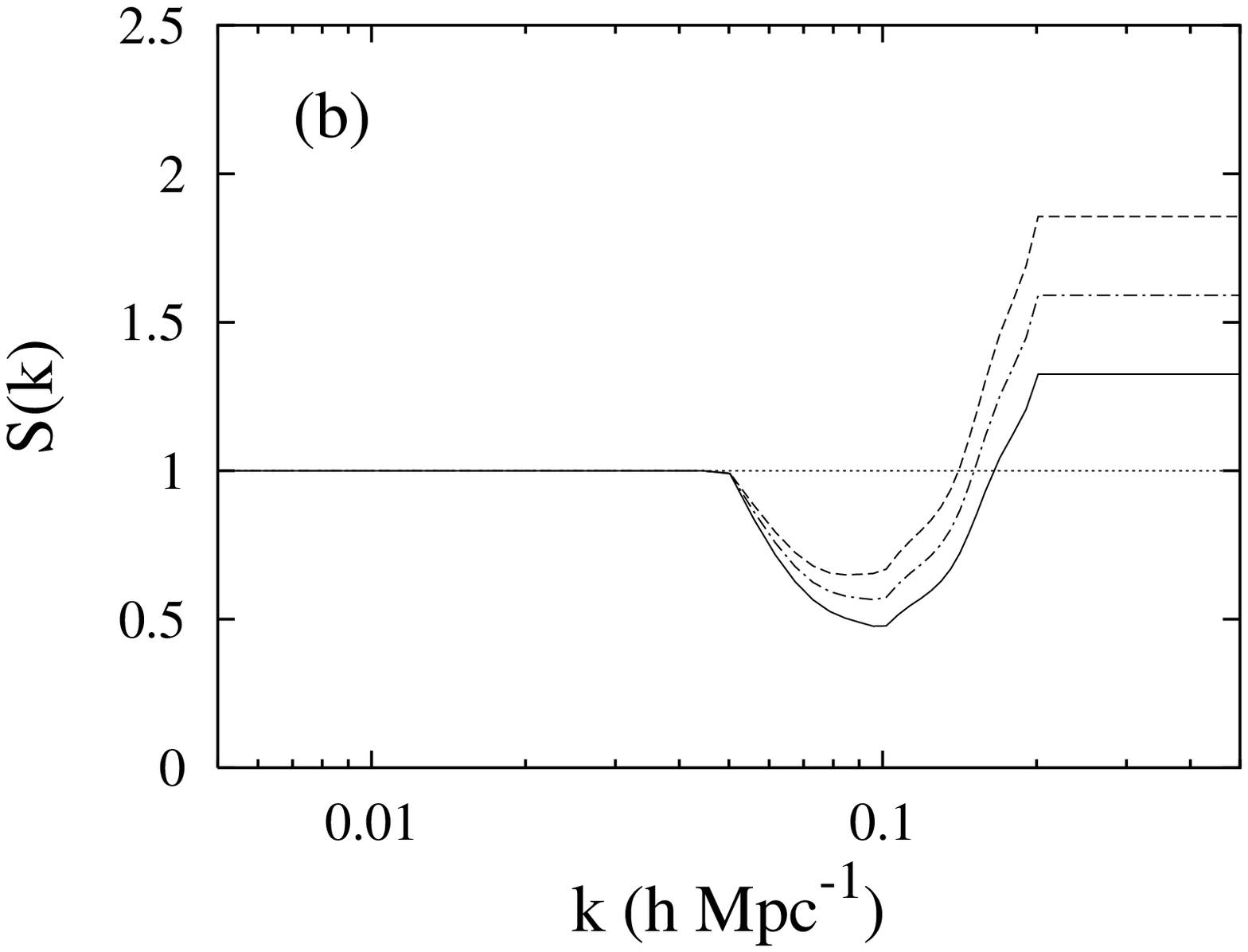,width=8cm}

\caption {(a) The power spectrum of density fluctuations in model
2 for $P(k_1)=2500h^{-3}$Mpc$^3$ (solid line), for
$P(k_1)=3000h^{-3}$Mpc$^3$ (dot-dashed line) and for
$P(k_1)=3500h^{-3}$Mpc$^3$ (dashed line). The dotted line shows
the power spectrum in the standard CDM model. (Here $h=0.65$.)
(b) The function $S(k)$ in the same models as in the panel (a).}
\end{figure}

Fig.~3a shows the power spectrum of density fluctuations in model 2
for different values of the parameter $P(k_1)$. In the models studied,
$P(k_1)=2500, 3000$ and $3500h^{-3}$Mpc$^3$. The normalized
Hubble constant $h=0.65$. In Fig.~3b, we show the function $S(k)$,
which describes the deviation of the initial power spectrum from the
scale-invariant form. The function $S(k)=1$ at wavenumbers
$k<0.05h$Mpc$^{-1}$, reaches the minimum at
$k_1= 0.1h$Mpc$^{-1}$ and then increases up to the wavenumber
$k_2=0.2h$Mpc$^{-1}$. At the minimum, $S(k_1)=0.48$ and
$S(k_1)=0.67$ for $P(k_1)=2500h^{-3}$ Mpc$^3$ and
$P(k_1)=3500h^{-3}$Mpc$^{3}$, respectively. For wavenumbers
$k>0.2h$Mpc$^{-1}$, we find that $S(k)=1.3$ and $S(k)=1.9$,
respectively. We investigated also the CDM model with $h=0.70$ assuming
that $P(k_1)=2500, 3000$ and $3500h^{-3}$ Mpc$^3$.
In this model, $S(k_1)=0.38$, $S(k_2)=1.0$ and $S(k_1)=0.52$,
$S(k_2)=1.4$ for the parameter $P(k_1)=2500h^{-3}$ Mpc$^3$ and
$P(k_1)=3500h^{-3}$Mpc$^{3}$, respectively.

Fig.~4 shows the variance, $\sigma^2(R)$, in the standard CDM model and
in model 2 for different values of the parameter $P(k_1)$. The variance
is given as a function of top-hat window radius $R$.
For the masses $M_{1.5}= 10^{14} h^{-1} M_{\odot}$ and
$M_{1.5}= 10^{15} h^{-1} M_{\odot}$, the window radius $R=5.8h^{-1}$
Mpc and $R=15.3h^{-1}$ Mpc, respectively (see eq. (4)).

\begin{figure}
\centering \leavevmode \psfig{file=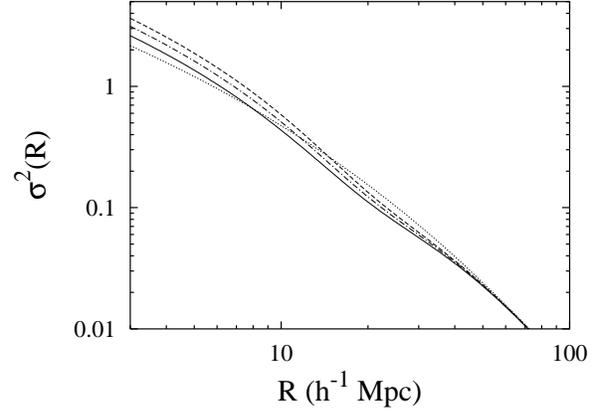,width=8cm} \caption
{The variance $\sigma^2(R)$ in the same models as in Fig.~3,
shown with the same line types.}
\end{figure}

Fig.~5 shows the cluster mass function as predicted in model 2 for the
parameter $P(k_1)=2500, 3000$ and $3500h^{-3}$ Mpc$^3$. The mass function
of clusters in the standard CDM model is also plotted. Fig.~5a shows the
results for $h=0.65$ and Fig.~5b for $h=0.70$. In model 2, the
mass function is steeper than that in the standard CDM model. At
the same value of $P(k_1)$, the cluster mass function for
$h=0.65$ and $h=0.70$ is similar. For comparison, we also show in
Fig.5 the observed mass function of clusters of galaxies derived
by BC and G98. In the models studied, the cluster mass function
is consistent with the observed data. If $h=0.65$, then for the
mass $M_{1.5}=4\cdot 10^{14} h^{-1} M_{\odot}$, the cluster
abundance $n(>M)=1.4 \cdot 10^{-6} h^3$Mpc$^{-3}$ and
$n(>M)=3.5 \cdot 10^{-6} h^3$Mpc$^{-3}$ for
$P(k_1)=2500h^{-3}$Mpc$^3$ and
$P(k_1)=3500h^{-3}$ Mpc$^3$, respectively. (If $h=0.70$,
$n(>M)=1.5 \cdot 10^{-6} h^3$Mpc$^{-3}$ and $n(>M)=3.7 \cdot
10^{-6} h^3$Mpc$^{-3}$, respectively.)

\begin{figure}
\centering \leavevmode \psfig{file=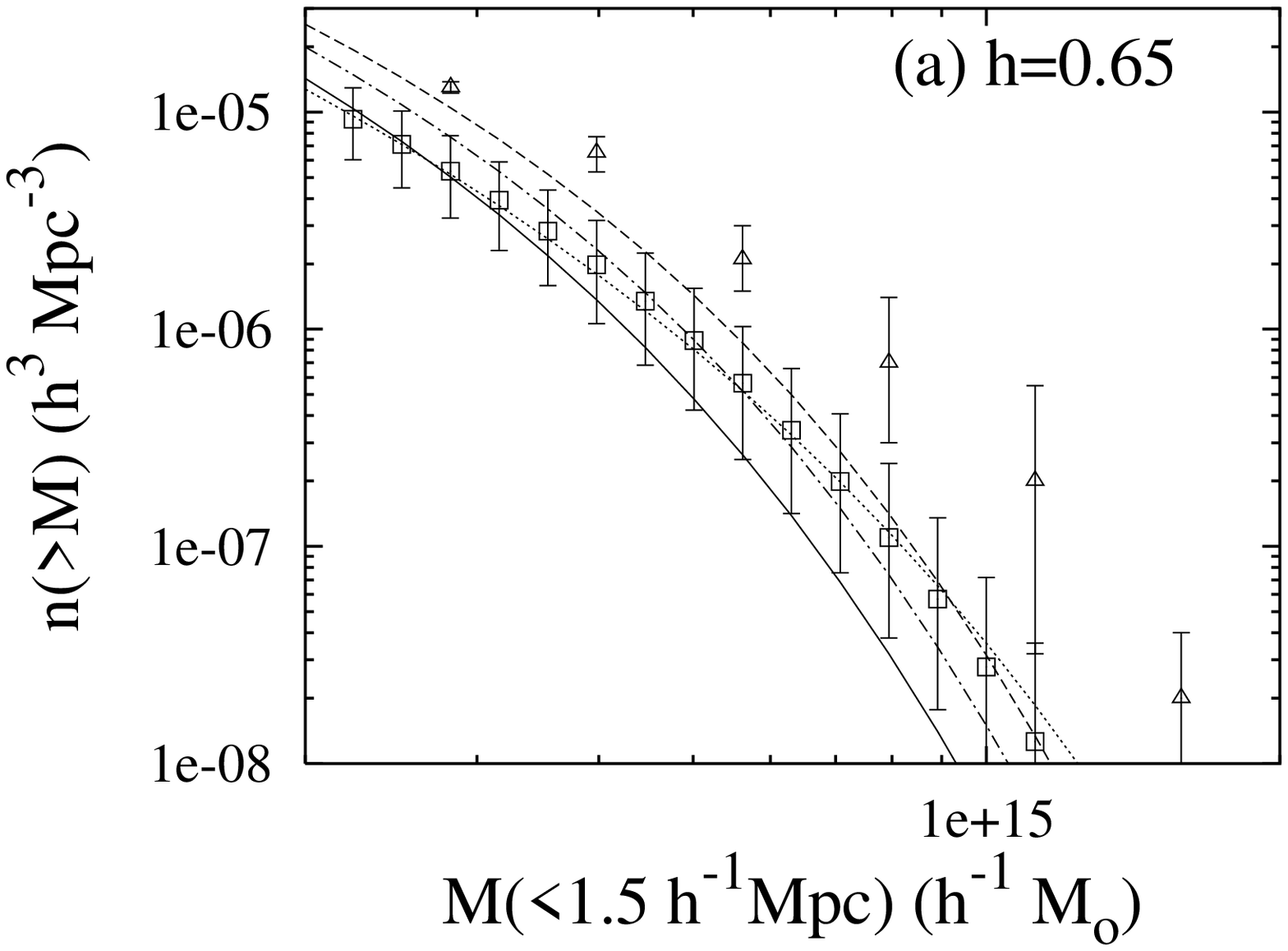,width=8cm}
\psfig{file=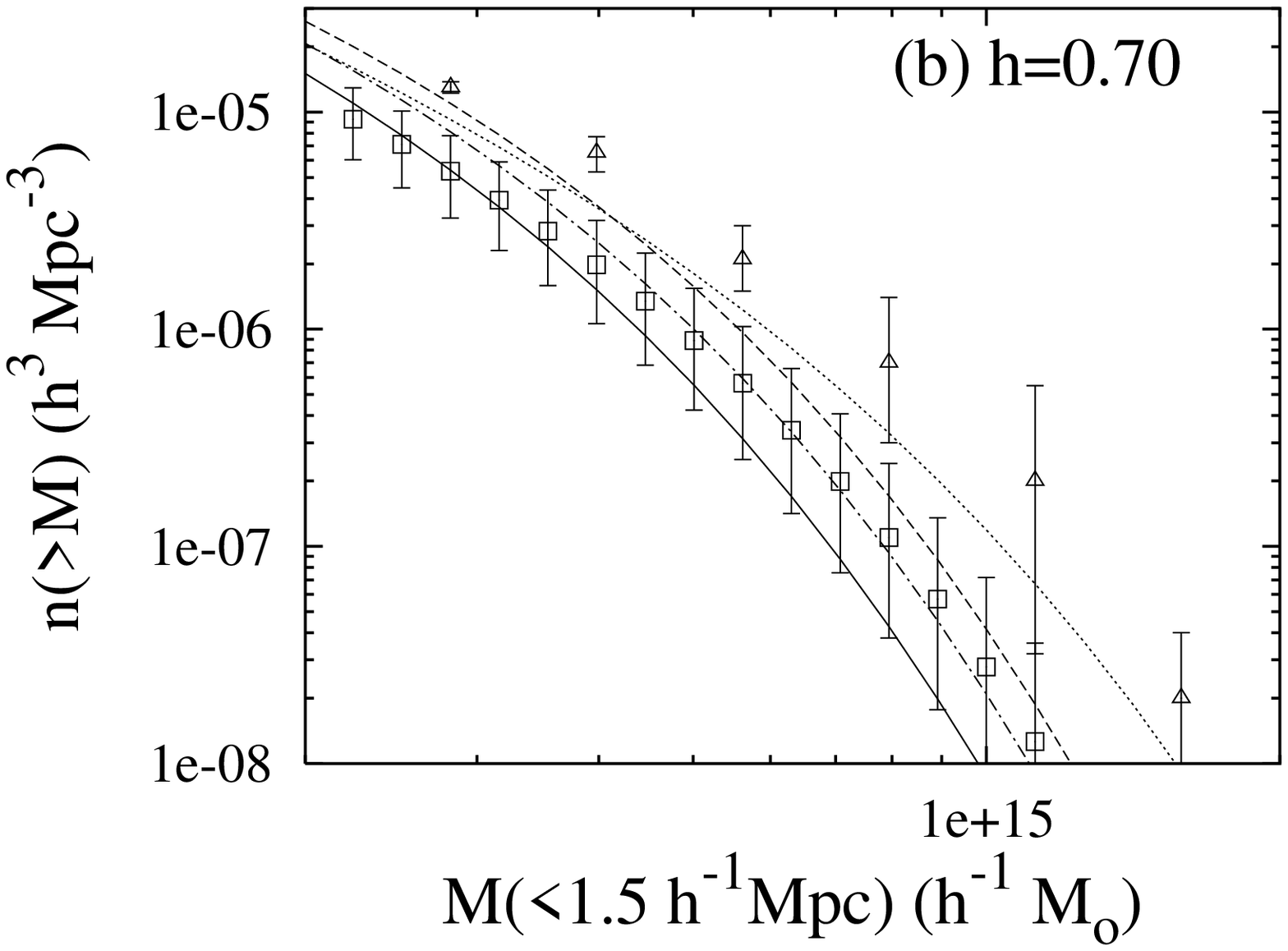,width=8cm}
\caption {The mass function of
clusters as predicted in model 2. (a) h=0.65. (b) h=0.70. The
lines are specified similarly to the Fig.~3. Squares and
triangles show the mass function of galaxy clusters derived by
Bahcall \& Cen (1993) and Girardi et al. (1998), respectively.}
\end{figure}

\begin{figure}
\centering \leavevmode \psfig{file=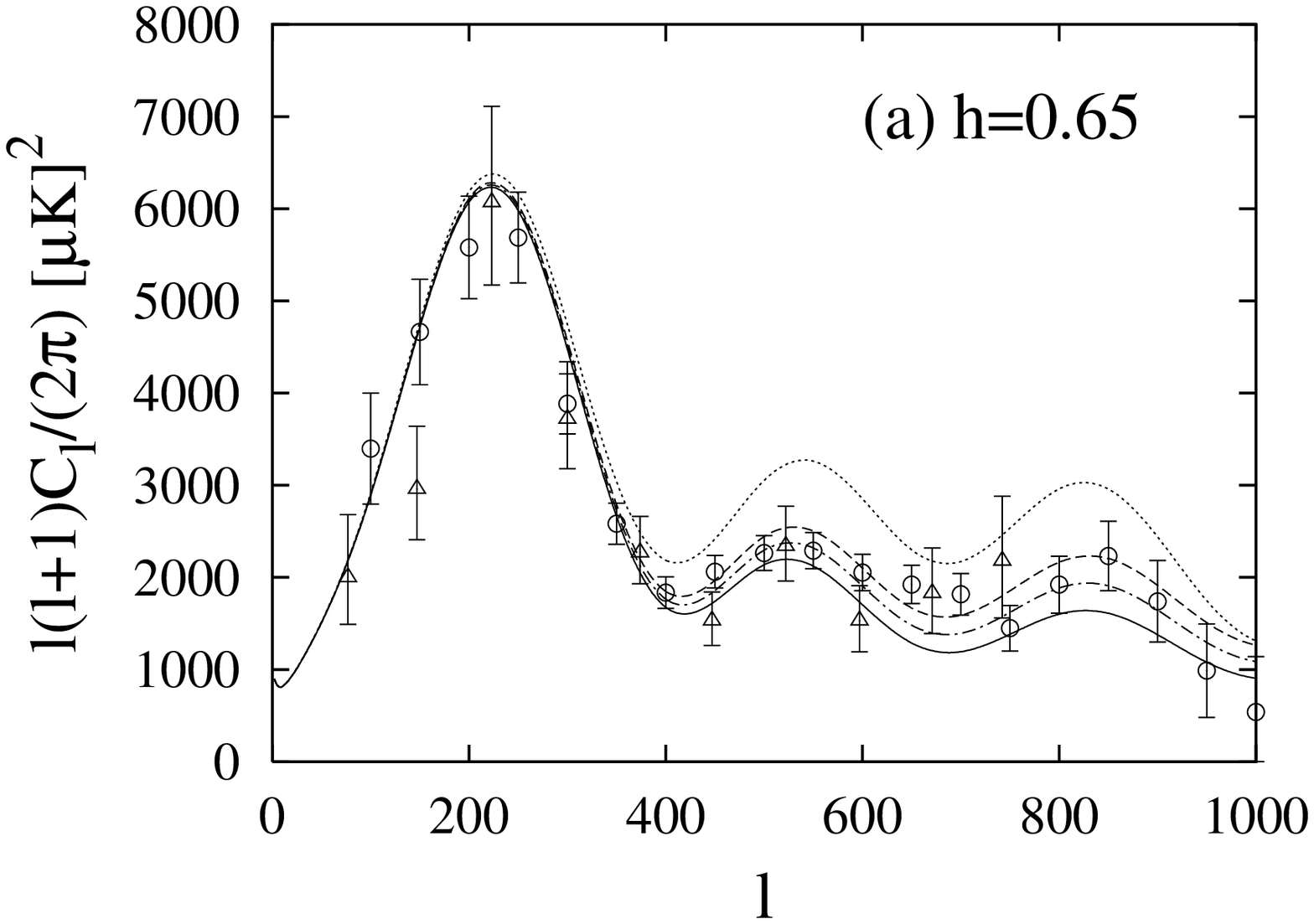,width=8cm}
\psfig{file=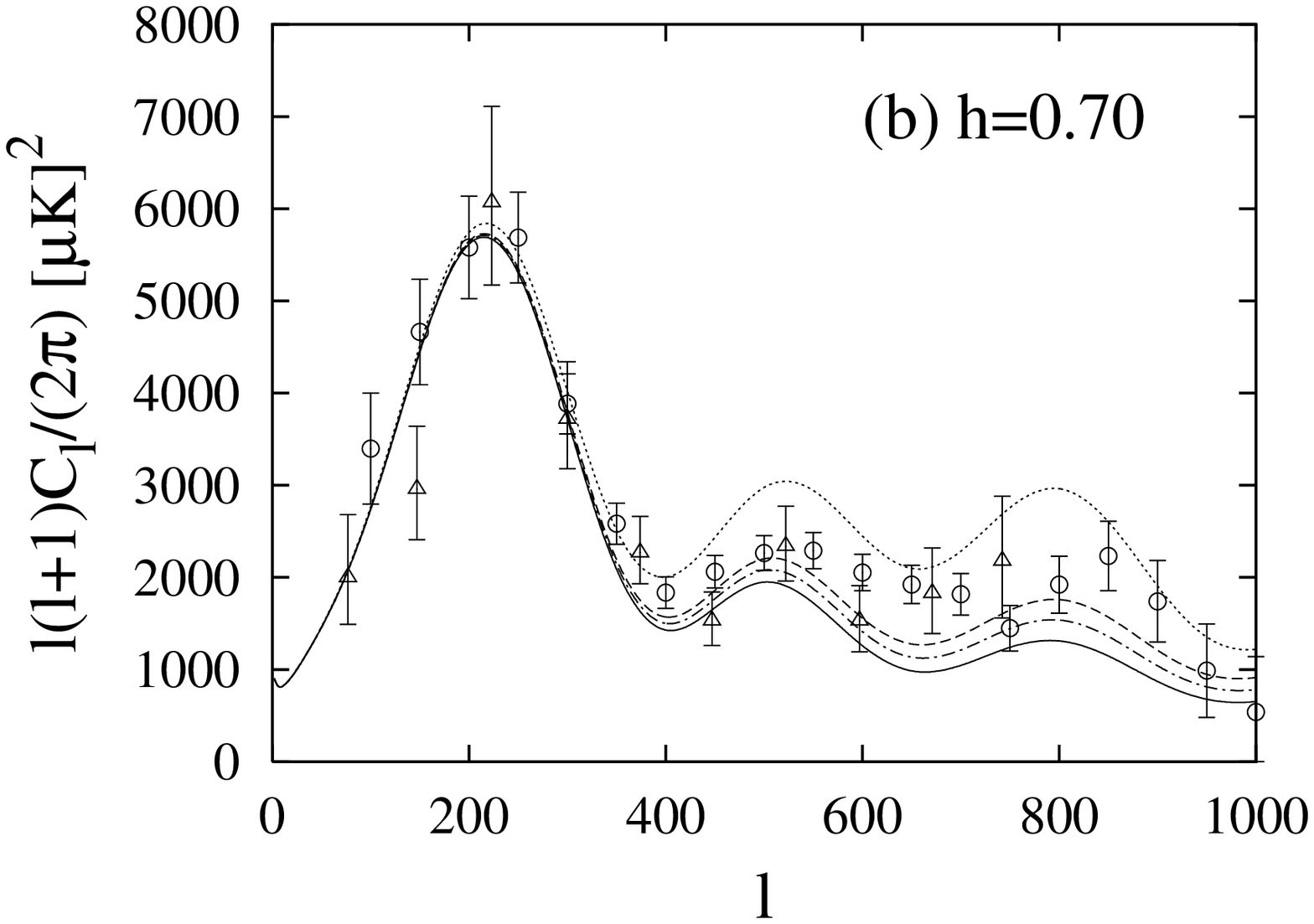,width=8cm} \caption {The power spectrum of
the CMB temperature in the same models as in Fig.~3, shown with
the same line types. (a) h=0.65. (b) h=0.70. The observational
data are from the Boomerang (circles) and Maxima-1 (triangles)
experiments.}
\end{figure}

Fig.~6 demonstrates the angular power spectrum of the CMB temperature
fluctuations as predicted in model 2. We also show the temperature power
spectrum in the standard CDM model with a scale-invariant power spectrum.
Fig.~6a demonstrates the results for $h=0.65$ and Fig.~6b for $h=0.70$.
In model 2, the amplitude of the temperature power spectrum at multipoles
$l>400$ is smaller than that predicted in the standard CDM model.
Fig.~6 also shows the CMB power spectrum derived from the Boomerang
(Netterfield et al. 2001) and from the Maxima-1 (Hanany et al. 2000)
experiments. In the models studied, the temperature power spectrum is
consistent with the observed temperature power spectrum.

Therefore, in the model with a bump in the power spectrum of density
fluctuations at the scale of superclusters,
the mass function of clusters and the temperature power spectrum are
in good agreement with the observed data.

\sec{PECULIAR VELOCITIES}

The observed rms peculiar velocity of galaxy clusters has been
studied in several papers (e.g. Bahcall, Gramann \& Cen 1994,
Bahcall \& Oh 1996, Borgani et al. 1997, Watkins 1997, Dale et
al. 1999). Watkins (1997) developed a likelihood method for
estimating the rms peculiar velocity of clusters from
line-of-sight velocity measurements and their associated errors.
This method was applied to two observed samples of cluster
peculiar velocities: a sample known as the SCI sample (Giovanelli
et al. 1997) and a subsample of the Mark III catalog (Willick et
al. 1997).  Watkins (1997) found that the rms one-dimensional
cluster peculiar velocity is $265^{+106}_{-75}$ km s$^{-1}$, which
corresponds to the three-dimensional rms velocity
$459^{+184}_{-130}$ km s$^{-1}$. Dale et al. (1999) obtained
Tully-Fisher peculiar velocities for 52 Abell clusters distributed
throughout the sky between $\sim 50$ and $200h^{-1}$Mpc. They
found that the rms one-dimensional cluster peculiar velocity is
$341\pm 93$km s$^{-1}$, which corresponds to the
three-dimensional rms velocity $591 \pm 161$ km s$^{-1}$.

To investigate peculiar velocities of clusters in our models,
we use the linear theory predictions for peculiar velocities of
peaks in the Gaussian field. The linear rms velocity fluctuation
on a given scale $R$ at the present epoch can be expressed as
$$
\sigma_v(R)=H_0 f(\Omega_0) \sigma_{-1} (R),
\eqno(6)
$$
where $f(\Omega_0) \approx \Omega_0^{0.56}$ is the linear
velocity growth factor in the flat models and  $\sigma_j$ is defined for
any integer $j$ by
$$
\sigma_j^2={1 \over 2\pi^2} \int P(k) W^2(kR) k^{2j+2} dk.
\eqno(7)
$$
Bardeen et al. (1986) showed that
the rms peculiar velocity at peaks of the smoothed density field differs
systematically from $\sigma_v(R)$, and can be expressed as
$$
\sigma_p(R)=\sigma_v(R) \sqrt{1 - \sigma_0^4/\sigma_1^2 \sigma_{-1}^2} .
\eqno(8)
$$

SG examined the linear theory predictions for the peculiar velocities of
peaks and compared these to the peculiar velocities of clusters in N-body
simulations. The N-body clusters were determined as peaks of the density
field smoothed on the scale $R \sim 1.5h^{-1}$ Mpc. The numerical results
showed that the rms peculiar velocity of small clusters is similar to
the linear theory expectations, while the rms peculiar velocity of rich
clusters is higher than that predicted in the linear theory.
The rms peculiar velocity of clusters with a mean cluster separation
$d_{cl} = 30h^{-1}$ Mpc was $\sim 18$ per cent higher than that predicted
by the linear theory. We assume that the observed cluster samples studied by
Watkins (1997) and Dale et al. (1999) correspond to the model clusters
with a separation $d_{cl}\sim 30h^{-1}$ Mpc
($n_{cl} \sim 3.7 \cdot 10^{-5} h^3$ Mpc$^{-3}$)
and determine the rms peculiar velocity of the
clusters, $v_{cl}$, as
$$
v_{cl}=1.18 \, \sigma_p(R),
\eqno(9)
$$
where the radius $R=1.5h^{-1}$ Mpc.

\begin{table}
\caption{Peculiar velocities in the different models.}
\begin{tabular}{|c|c|c|c|}

\hline
 $h$ &    $P(k_1)$         &   $v_{cl}$ & $V_{60}$ \\
     & ($h^{-3}$Mpc$^{-3}$)& (km s$^{-1}$) & (km s$^{-1}$) \\
\hline
           0.65 & 2500 & 557 & 273  \\
                & 3000 & 578 & 273  \\
                & 3500 & 598 & 273  \\
           0.70 & 2500 & 579 & 284  \\
                & 3000 & 599 & 284  \\
                & 3500 & 619 & 284  \\
\hline
\end{tabular}
\label{table}
\end{table}

In the standard CDM model with $\Omega_m=0.3$, we found that
$v_{cl}=582$ km s$^{-1}$ and $v_{cl}=635$ km s$^{-1}$ for
$h=0.65$ and $h=0.70$, respectively. Table~1 lists the rms peculiar
velocity of clusters, $v_{cl}$, in our models. The rms peculiar velocity
of clusters is $\sim 555-620$ km s$^{-1}$, which is consistent with the
observed rms peculiar velocity of clusters derived by Watkins (1997) and
Dale et al. (1999).

We also studied the rms bulk velocity for a radius
$r=60h^{-1}$Mpc, $V_{60}$. The rms bulk velocity was determined by
using equation (6). In the standard $\Omega_m=0.3$ model,
$V_{60}=273$ km s$^{-1}$ and $V_{60}=285$ km s$^{-1}$ for
$h=0.65$ and $h=0.70$, respectively. Table~1 lists the rms bulk
velocity, $V_{60}$, in our models. The bulk velocity is similar
to that in the standard model. The observed bulk velocities are
determined in a sphere centred on the Local Group and represent a
single measurement of the bulk flow on large scales. The observed
bulk velocity derived from the Mark III catalogue of peculiar
velocities for $r=60h^{-1}$ Mpc is $370\pm 110$ km s$^{-1}$
(Kolatt \& Dekel 1997). Giovanelli et al. (1998) studied the bulk
velocity in the SCI sample and estimated that the bulk flow of a
sphere of radius $r=60h^{-1}$ Mpc is between $140$ and $320$ km
s$^{-1}$. In the models studied, the rms bulk velocity is $\sim
275-285$ km s$^{-1}$, which is consistent with the observed data.

\sec{DISCUSSION AND SUMMARY}

In this paper we have examined a CDM model, where the power spectrum
contains a specific feature (bump) at the wavenumbers
$k\sim 0.1-0.2h^{-1}$ Mpc,
which correspond to the scale of superclusters of galaxies.
We studied a flat cosmological model with the density
parameter $\Omega_m=0.3$ and the normalized Hubble constant $h=0.65$ and
$h=0.70$. The baryon density was assumed to be consistent with the
standard big-bang nucleosynthesis (BBN) value. We investigated the mass
function of clusters and the angular power
spectrum of the CMB temperature fluctuations, assuming different values
of the spectral
parameter $P(k_1)$ that determines the amplitude of the power spectrum
for the bump. We found that the cluster mass function and the CMB
power spectrum are in good agreement with the observed data if the spectral
parameter $P(k_1)$ is in the range $P(k_1)=2500-3500h^{-3}$Mpc$^3$.
We also investigated the rms peculiar velocity of clusters and the rms
bulk velocity for a radius $r=60h^{-1}$ Mpc. In the models studied, the rms
peculiar velocity of clusters is $\sim 555-620$ km s$^{-1}$ and the rms
bulk velocity is $\sim 275-285$ km s$^{-1}$, which are consistent with the
observed data.

Therefore, in many aspects the CDM model, where the power spectrum
contains a feature at the scale of superclusters of galaxies, fits the
observed data. This model predicts that there is a bump in the correlation
function of clusters at separations $r\sim 20-35h^{-1}$Mpc
(Suhhonenko \& Gramann 1999). In the future, accurate
measurements of the cluster correlation function at these distances can
serve as a discriminating test for this model.

What could be the origin of the primordial feature examined in
this paper (see Fig.~3b)? The standard inflationary prediction
concerning the initial power spectrum of density fluctuations is
a simple power law: $P_{in} \sim k^n$. However, for more than ten
years, there has been some interest in models called
broken-scale-invariant (BSI), predicting deviations from a
power-law. These models generally involve, in addition to the
usual inflation field, other (effective) fields, driving
successive stages of inflation or just triggering a phase
transition (e.g. Starobinsky 1985; Kofman, Linde \& Starobinsky
1985; Kofman \& Pogosyan 1988; Gottl\"ober, M\"uller \&
Starobinsky 1991). Recently, there have been efforts to implement
double/multiple inflation in realistic supersymmetric contexts
(Lesgourgues 1998). There has also been an attempt to generate
spectral features via resonant production of particles during
inflation (Chung et al. 2000). Clearly, further work is needed to explain
the primordial feature introduced in this paper in the context of
different inflationary models.

\sec*{ACKNOWLEDGEMENTS}

We thank J. Einasto, M. Einasto and E. Saar for useful discussions.
This work has been supported by the ESF grant 3601.

\vfill

\end{document}